\documentclass[english]{article}
\usepackage[nottoc, numbib]{tocbibind}

\usepackage{hyperref}
\usepackage[letterpaper]{geometry}
\usepackage{amsmath}
\usepackage{amssymb}
\usepackage{algorithm}
\usepackage{algpseudocode}
\usepackage{bm}
\usepackage{graphicx}
\usepackage{float}
\usepackage{array}
\usepackage{tikz}
\usepackage{enumerate}
\usepackage{booktabs}
\usepackage{tabularx}
\usepackage{caption} 
\usepackage{authblk}
\usepackage{parskip}
\usepackage[round, sort, numbers, square]{natbib}
\usepackage{soul}
\usepackage{multirow}
\usepackage{wrapfig}
\usepackage[shortlabels]{enumitem}
\usepackage [english]{babel}
\usepackage [autostyle, english = american]{csquotes}

\newcommand{\R}{\mathbb{R}}

\title{Curiosity as filling, compressing, and reconfiguring knowledge networks}

\author[a]{Shubhankar P. Patankar}
\author[b]{Dale Zhou}
\author[c,d]{Christopher W. Lynn}
\author[a]{Jason Z. Kim}
\author[e]{Mathieu Ouellet}
\author[b]{Harang Ju}
\author[f]{Perry Zurn}
\author[a,g,h]{David M. Lydon-Staley}
\author[a,e,i,j,k,l,*]{Dani S. Bassett} 

\affil[a]{Department of Bioengineering, School of Engineering and Applied Science, University of Pennsylvania, Philadelphia, PA 19104 USA}
\affil[b]{Neuroscience Graduate Group, Perelman School of Medicine, University of Pennsylvania, PA 19104 USA}
\affil[c]{Initiative for the Theoretical Sciences, Graduate Center, City University of New York, New York, NY 10016 USA}
\affil[d]{Joseph Henry Laboratories of Physics, Princeton University, Princeton, NJ 08544 USA}
\affil[e]{Department of Electrical and Systems Engineering, School of Engineering and Applied Science, University of Pennsylvania, Philadelphia, PA 19104 USA}
\affil[f]{Department of Philosophy, American University, Washington, DC 20016 USA}
\affil[g]{Annenberg School for Communication, University of Pennsylvania, Philadelphia, PA 19104 USA}
\affil[h]{Leonard Davis Institute of Health Economics, University of Pennsylvania, Philadelphia, PA 19104 USA}
\affil[i]{Department of Psychiatry, Perelman School of Medicine, University of Pennsylvania, Philadelphia, PA 19104 USA}
\affil[j]{Department of Neurology, Perelman School of Medicine, University of Pennsylvania, Philadelphia, PA 19104 USA}
\affil[k]{Department of Physics and Astronomy, College of Arts and Sciences, University of Pennsylvania, Philadelphia, PA 19104 USA}
\affil[l]{Santa Fe Institute, Santa Fe, NM 87501 USA}
\affil[*]{To whom correspondence should be addressed: dsb@seas.upenn.edu}

\begin{document}
\maketitle


\newpage
\begin{abstract}

Curiosity is an internally motivated search for information. It is enduring and open-ended, and may have evolved to help us build accurate mental representations of our ever-changing environments. Due to the significant role that curiosity plays in our lives, several theoretical constructs, such as the information gap theory and compression progress theory, have sought to explain how we engage in its practice. According to the former, curiosity is the drive to acquire information that is missing from our understanding of the world. According to the latter, curiosity is the drive to construct an increasingly parsimonious mental model of the world. To complement the densification processes inherent to these two theories, we propose the conformational change theory, wherein we posit that the practice of curiosity results in mental models with marked conceptual flexibility. To validate these three theories, we must overcome the fundamental challenge of constructing formal models of mental representations of knowledge. Here, we address that challenge by formalizing curiosity as the process of building a growing knowledge network. We then quantitatively investigate information gap theory, compression progress theory, and the conformational change theory of curiosity. In knowledge networks, gaps can be identified as topological cavities, compression progress can be quantified using network compressibility, and flexibility can be measured as the number of conformational degrees of freedom. We leverage data acquired from the online encyclopedia Wikipedia to determine the degree to which each theory explains the growth of knowledge networks built by individuals and by collectives. Our findings lend support to a pluralistic view of curiosity, wherein intrinsically motivated information acquisition fills knowledge gaps and simultaneously leads to increasingly compressible and flexible knowledge networks. Across individuals and collectives, we determine the contexts in which each theoretical account may be explanatory, thereby clarifying their complementary and distinct explanations of curiosity. Our findings offer a novel network theoretical perspective on intrinsically motivated information acquisition that may harmonize with or compel an expansion of the traditional taxonomy of curiosity.

\end{abstract}

\newpage
\section{Introduction}

Humans must manage uncertainty and embrace change to thrive in a complex and dynamic environment \citep{Gottlieb_2013}. To this end, we continually consume information to construct and maintain accurate mental models of the world \cite{valadao2015examining,johnson2010mental}. Information-seeking behavior may be driven by a variety of intrinsic and extrinsic factors. Arising from the latter, information acquisition is an intermediate step towards attaining a specific goal---such as increased wealth or social recognition---that is ultimately rewarding \cite{dweck1986motivational}. By contrast, the intrinsic motivation to seek information is commonly conceptualized as curiosity \citep{Loewenstein_1994, Gottlieb_2013, Kidd_2015}. \textit{In silico} work suggests that curiosity may have evolved to maximize long-term evolutionary fitness in rapidly changing environments \citep{Satinder_2010}. Additionally, studies have shown that humans are driven to know, even when information is costly to obtain \cite{hsee2016pandora,clark2021smokers} and may have no immediate tangible utility \citep{Bennett_2016, Brydevall_2018}. Curiosity-driven information gathering is, therefore, inherently rewarding. 

Given the significant role that it plays in our daily behavior and decision making, several theories have sought to explain how individuals practice curiosity. The \textit{information gap theory} views curiosity as the drive to obtain information that is missing from a mental model of the world \citep{Loewenstein_1994}. In this account, perception of a gap in one's knowledge of the world creates an aversive state of uncertainty that, in turn, motivates a search for information to close the gap \citep{Daddaoua_2016}. In the complementary perspective of \textit{compression progress theory}, curiosity is the drive to obtain information that improves the compression of a mental model and thereby lowers its cost of representation \citep{Schmidhuber_2008, Zhou_2020}. Both theories provide several important explanations for curiosity-driven information-seeking behavior. On the one hand, curiosity that explicitly modulates uncertainty levels by opening or closing information gaps as needed can facilitate lifelong learning. On the other hand, compression progress enables the extraction of essential latent structures of knowledge, offering greater capacity for generalization \citep{Tenenbaum_2011, Collins_2017}. Low-cost representations also allow for greater functional capacity by freeing cognitive resources that would otherwise be dedicated to information storage and retrieval \citep{Wolff_2019}. However, theoretical constructs for curiosity, such as the information gap theory and compression progress theory, are difficult to validate quantitatively. A fundamental challenge is constructing formal models of mental representations of knowledge. In the absence of such models (and the conceptual frameworks that accompany them), it is unclear how to translate theoretical concepts such as knowledge gaps and compression progress to well-defined variables that can be measured in experiments. 

\begin{figure}[htbp] 
    \begin{center}
    \includegraphics[width=\textwidth]{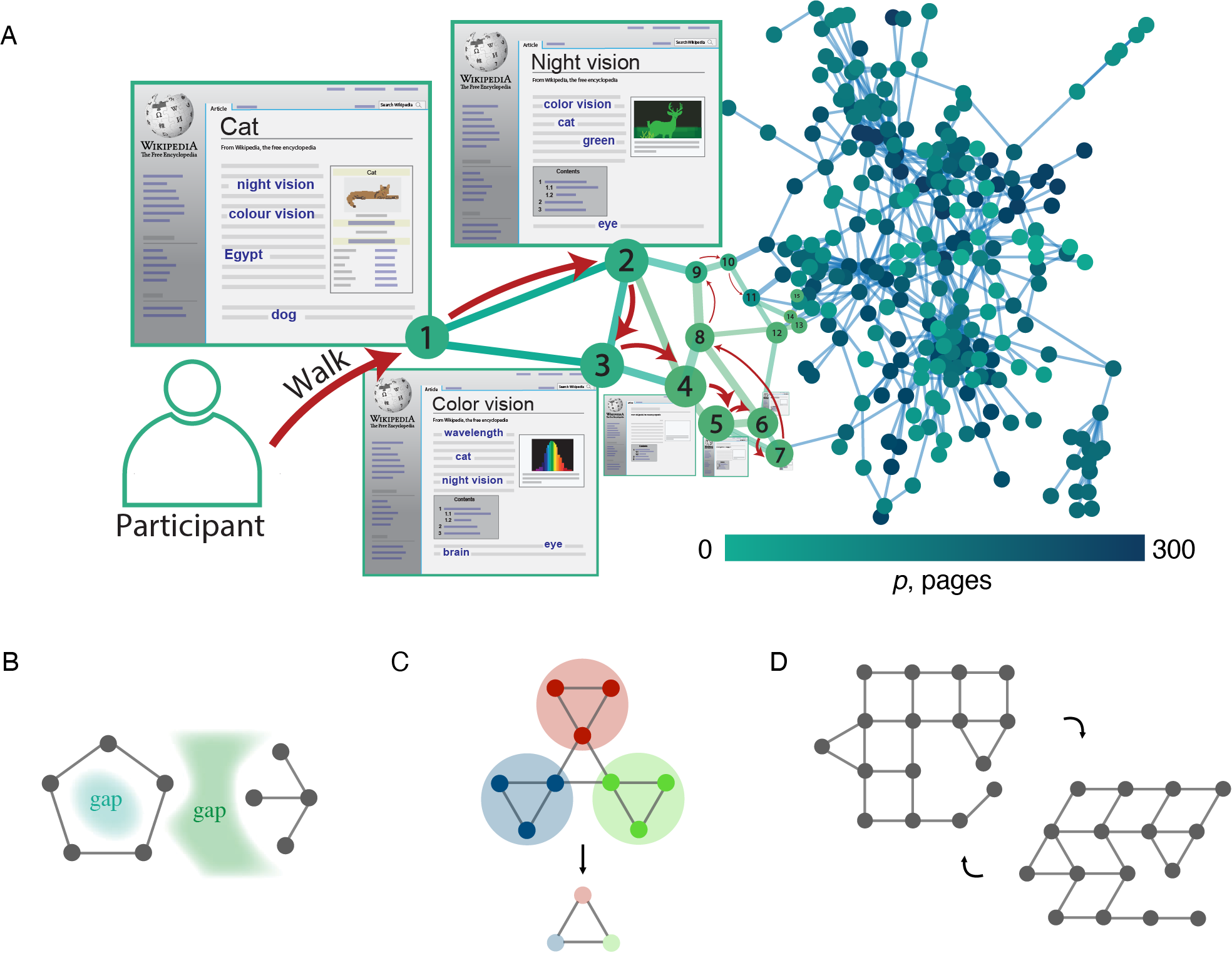}
    \end{center}
    \caption{\label{fig:schematic} \textbf{Connectional approach to curiosity.} \emph{(A)} A participant constructs a growing knowledge network through curiosity-driven self-directed exploration of Wikipedia, a vast networked landscape of information. Nodes represent unique Wikipedia pages. Edges represent hyperlinks between nodes. Nodes are colored to denote the order in which they are visited. \emph{(B)} Gaps in a knowledge network can be formalized using algebraic topology and tracked in several topological dimensions. The green and blue gaps represent a $0$-dimensional and $1$-dimensional cavity, respectively. \emph{(C)} Compression progress aims to construct internal representations of the world that are both storage efficient and generalizable. In a knowledge network, all concepts that belong to the same cluster can be represented parsimoniously at a higher level of abstraction using their cluster identity. The unclustered network has $9$ nodes and $12$ edges, while the clustered network only has $3$ nodes and $3$ edges. \emph{(D)} A mechanical network can possess several spatial configurations, any of which can be arrived at from any of the others through a series of conformational changes. We formalize and measure knowledge network flexibility as the number of available conformational degrees of freedom.}
\end{figure}

One such model that has shown promise is a network model where knowledge is composed of discrete and yet interconnected concepts \citep{Chrastil_2014, chrastil2015active, Schapiro_2016, Warren_2017, ericson2020probing, Peer_2021, Stiso_2022}. In graph learning studies, volunteers are shown sequences of images on a screen, where, unbeknownst to the volunteers, each image corresponds to a node in an underlying network \cite{Lynn_2020}. Based solely on observed transitions, and despite being unaware of the underlying network's structure, participants successfully infer statistical regularities from the temporal order in which images appear \citep{Schapiro_2013, Schapiro_2016, Garvert_2017,kahn2018network,tompson2019individual}. Crucially, the structure of the pre-defined experimental graph can be recovered from neural activity by decoding simultaneously acquired functional magnetic resonance imaging (fMRI) data \citep{Garvert_2017,tompson2020functional}. The sequential manner in which stimuli are presented in graph learning tasks can be conceived of as a walk prescribed by the experimenter in a limited knowledge space of objects, images, concepts, or movements. Curiosity, too, can be conceived of as a walk, but one that is largely self-directed and purposeful across the vast landscape of knowledge. To evaluate curious walks, recent work gathered browsing histories from individuals who freely explored the online encyclopedia Wikipedia. Structural features of the knowledge networks that participants walked upon (Fig. \ref{fig:schematic}A) were found to be associated with curiosity, as measured by an independent index of participants' sensitivity to information deprivation \citep{LydonStaley_2021}.

Here, we leverage this framework and cast curiosity as a network building process. This approach allows us to take qualitative explanations for information-seeking behavior, such as the information gap theory and compression progress theory, and operationalize them in quantitative statistics. Information gap theory posits that humans add information to regulate uncertainty by filling gaps \cite{Loewenstein_1994, Gottlieb_2013, Kidd_2015}. This theory can be operationalized by treating gaps in networks as topological cavities, and by tracking their evolution using techniques from applied algebraic topology \citep{Hatcher_2002, Ghrist_2007, Bianconi_2021} (Fig. \ref{fig:schematic}B). In contrast, compression progress theory posits that humans subtract or discard information \cite{Schmidhuber_2008, Zhou_2020, Lynn_2021} due to limited cognitive capacity \cite{shiffrin1977controlled, Zhou_2020, Zhou_2021}. Compressing a network while maintaining meaningful latent structure requires that we discard some irrelevant information while maintaining important information about past experiences and present priorities \cite{Lynn_2021, momennejad2020learning, Zhou_2020, lynn2020abstract}. This theory can be operationalized by measuring the compressibility of a network, an information-theoretic quantity that captures the ability of a network to be compressed \citep{Lynn_2021} (Fig. \ref{fig:schematic}C). Via the network operationalization of these two theories, we come to see that curiosity is marked as a process by which networks of knowledge densify and simplify, raising the question of what alternative process might drive them to sprawl and become complex.

To address this question, we expand beyond historical accounts to operationalize our own conformational change theory of curiosity. The conformational change theory suggests that information-seeking behavior results in the creation of expansive knowledge networks \citep{Zurn_2021} embedded in a conceptual geometry (Fig. \ref{fig:schematic}D). The notion of a conceptual geometry is motivated by prior studies of neural population geometry and the fact that information can be embedded and processed in locally Euclidean geometric representations to solve complex tasks \citep{Chung_2021}. The geometry provides a key affordance for curiosity---conceptual flexibility---as the knowledge network can mechanically conform into different shapes. While some concepts are separated from other concepts by fixed distances of shared-versus-unshared meaning, other concept pairs can move closer together or farther apart as inter-concept relations shorten or lengthen depending on time and context \citep{Kim_2019}. This flexibility allows us to draw from past experience, cohere the past with newly learned information, monitor conflict, and respond appropriately in different contexts \cite{Tenenbaum_2011, botvinick2015motivation, karuza2016local}; it may also subserve the unexpected conceptual combinations that accompany imaginative thought and support serendipitous discoveries \cite{mcallister2012thought,copeland2019serendipity}. Mechanically akin to conformational change in proteins, the flexible reshaping of the knowledge network can only occur if concepts are sparsely connected; densely connected linkage networks embedded in a Euclidean geometry are rigid. Hence, the conformational change theory of curiosity posits a drive for conceptual flexibility that leads networks to sprawl and become complex.

As is now evident, each of the three theories is motivated by a distinct and uniquely important psychological drive: to reduce uncertainty by learning a missing piece of information, to discover latent patterns by distilling fundamental epistemic elements, and to reshape information by flexibly reconfiguring knowledge networks. Here, we test each theory through parallel analyses of the growth of individual and collective knowledge networks derived from Wikipedia. At the individual scale, we construct knowledge networks for $149$ individuals using their Wikipedia browsing histories \citep{LydonStaley_2021} (Fig. \ref{fig:schematic}A). At the collective scale, we extract Wikipedia networks to assess knowledge growth in $30$ disciplines such as calculus, economics, and linguistics \citep{Ju_2020}. We treat Wikipedia pages as nodes in both sets of networks and add edges between them according to the presence of hyperlinks between pages. For the data on individuals, we specify network growth using the order in which individuals visit pages; for the data on collectives, we use the years in which different concepts originate. To model the random growth of knowledge in both data sets, we create $25$ degree-preserving edge-rewired versions of each network. We test the predictions of the three theories by comparing measurements of relevant features from empirically observed knowledge networks to those from the related null networks. First, considering the information gap theory, we expect to find fewer-than-chance topological cavities in growing empirical knowledge networks due to people's hypothesized drive to close knowledge gaps when they are perceived. Considering compression progress theory, we hypothesize that growing knowledge networks will exhibit greater-than-chance compressibility due to people's hypothesized drive to distill fundamental epistemic elements \citep{Zhou_2020}. Third, considering conformational change theory \citep{Zurn_2021}, we hypothesize that knowledge networks will possess greater-than-chance capacity for conformational changes due to people's hypothesized drive for conceptual flexibility. In testing these hypotheses, we demonstrate the utility of the network approach in quantitatively validating existing theoretical constructs of curiosity as well as in formulating new ones.

\section{Results}

\subsection{Network growth formalism} \label{sec:network_growth}

Before testing the predictions of the three theories, we clarify the network formalism upon which they are operationalized. Consider a graph $\mathcal{G} = (\mathcal{V}, \mathcal{E})$ with node set $\mathcal{V}$ and edge set $\mathcal{E} \subseteq \mathcal{V} \times \mathcal{V}$. We define a growing knowledge network with the tuple $(\mathcal{G},\; s)$, where $s$ denotes a map $s: \mathcal{V} \rightarrow \mathbb{N}$ that specifies the rank order in which nodes are added to the network. For networks built by individuals, $s$ is determined by the order in which Wikipedia pages are first visited. For collective networks, $s$ is determined by the years in which different concepts originate. With $N$ nodes in a network, we construct a sequence of graphs 
\begin{equation} \label{eq:graph_filtration}
    \mathcal{G}_{0} \subset \mathcal{G}_{1} \subset \cdots \subset \mathcal{G}_{N} = \mathcal{G},
\end{equation}
where $\mathcal{G}_p$ is a subgraph of $\mathcal{G}$ comprised of the first $p$ nodes in $s$ and all $q$ connections between them that exist in $\mathcal{E}$. Such a sequence---in which each element is a subset of the next---is an example of a \textit{filtration}. We index each subgraph in a filtration by the number of nodes in the network at that stage. We identify topological cavities, measure network compressibility, and compute conformational flexibility for all subgraphs in filtrations of individual and collective knowledge networks as well as in filtrations of related null model networks. We perform non-parametric permutation tests to examine differences between feature curves for empirical and null model data (see Sec. \ref{sec:statistical_testing} for details). Since networks in a given data set may have different sizes, we normalize filtration indices to span the range [$0$, $1$], and align values of interest to be defined on the same points before computing the mean for a feature-of-interest across all individuals or topics \citep{Christianson_2020}. For completeness, we also report results with unnormalized values in the Supplement.

\subsection{Information gap theory}

\begin{figure}[htbp]
	\begin{center}
	\includegraphics[width=\textwidth]{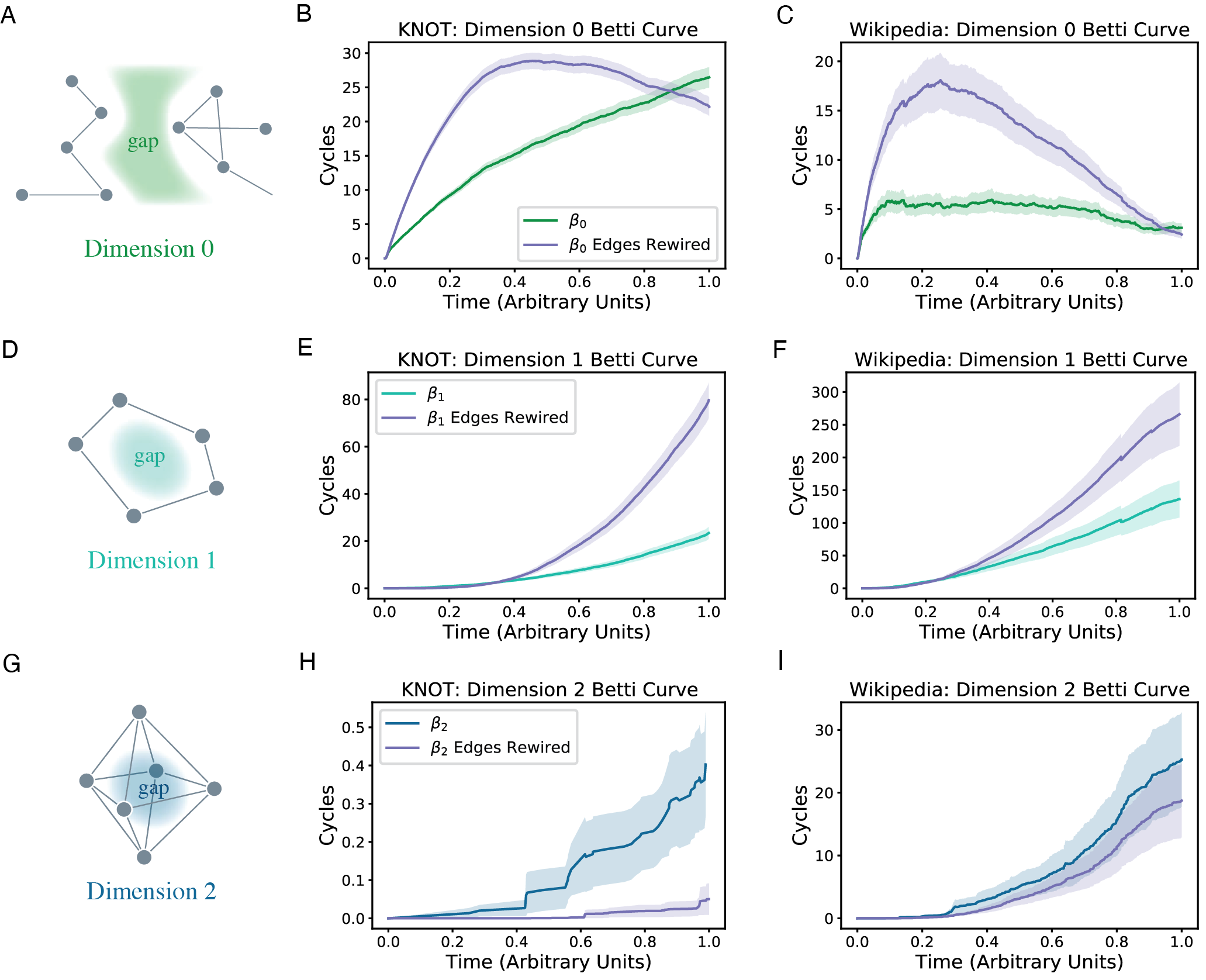}
	\end{center}
	\caption{\label{fig:IGT} \textbf{Probing information gaps as topological cavities in growing knowledge networks.} We operationalize information gaps as topological cavities (also referred to as \textit{cycles}) and track their evolution in growing individual and collective knowledge networks. We plot the number of cycles as a function of time. \emph{(A)} Topological cavities in dimension $0$, or $0$-cycles, represent disconnected network components. \emph{(B, C)} Individual and collective knowledge networks tend to possess fewer disconnected components than expected in edge-rewired null model networks. \emph{(D)} In dimension $1$, a topological cavity represents an enclosed loop formed by edges. \emph{(E, F)} Growing individual and collective knowledge networks tend to possess fewer loops than expected in edge-rewired null model networks. \emph{(G)} A topological cavity in dimension $2$ constitutes a void enclosed by $3$-cliques, or triangles of interconnected nodes. \emph{(H, I)} Growing individual and collective knowledge networks tend to possess more $2$-dimensional cavities than expected in edge-rewired null model networks. Shaded regions in panels \emph{B, C}, \emph{E, F}, and \emph{H, I} represent standard error. Purple curves denote the average number of cavities in edge-rewired null model networks.}
\end{figure}

The information gap theory posits that curiosity is the drive to collect units of knowledge that fill gaps in one's internal representation of the world \citep{Loewenstein_1994}. When we model internal representations as networks, the missing information can be usefully operationalized as topological cavities, which can be tracked in a principled manner using tools from applied algebraic topology (see Sec. \ref{sec:homology} for methodological details) \citep{Hatcher_2002, Ghrist_2007}. This operationalization follows prior work demonstrating that domains as diverse as language development in toddlers \citep{Sizemore_2018_1}, the introduction of characters in Dostoyevsky's novels \citep{Gholizadeh_2018}, and the presentation of concepts in linear algebra textbooks \citep{Christianson_2020} exhibit a systematic creation and closing of such cavities. By employing this approach, we can determine whether curiosity-driven exploration is motivated by a preference for gap closure. In a network, except for dimension $0$, a $k$-dimensional cavity, also known as a $k$-cycle, is identified as an empty enclosure formed from $(k+1)$-cliques, where cliques are defined as all-to-all connected subgraphs of $k+1$ nodes. The $k$-th Betti curve records the number of $k$-cycles present at each stage of a network's growth. Cycles of dimension $0$ represent disconnected network components (Fig. \ref{fig:IGT}A), whereas those of dimensions $1$ and $2$ represent loop-like holes (Fig. \ref{fig:IGT}D) and pocket-like voids (Fig. \ref{fig:IGT}G), respectively. 

Considering information gap theory, we hypothesized that empirical knowledge networks would contain fewer cavities than topologically similar edge-rewired null model networks. To test this hypothesis, we compute persistent homology for filtrations of individual and collective knowledge networks in dimensions $0$, $1$, and $2$. We find that the number of $0$-cycles, or disconnected network components, increases as individual knowledge networks grow, and does so at a steeper rate in null networks than in empirical networks (Fig. \ref{fig:IGT}B). For collective knowledge networks, we find that the number of disconnected components first increases and then decreases both in the empirical and in the null model networks, albeit with significantly different peak values (Fig. \ref{fig:IGT}C). In both data sets, for a significant duration of growth, Betti curves for observed networks are lower than those for null model networks. In dimensions $1$ and $2$, we find that the number of cycles increases as individual and collective knowledge networks grow (Fig. \ref{fig:IGT}D-I). This temporal trajectory could arise from the fact that filling gaps by forging new connections can open new gaps, making it prohibitively difficult to track (and fill) gaps among an increasingly large number of items. In support of information gap theory, the rate at which $1$-cycles increase is lower for the empirical networks than for the null networks (Fig. \ref{fig:IGT}E,F). In contrast to information gap theory, the rate at which $2$-cycles increase is higher in the empirical networks than in the null networks (Fig. \ref{fig:IGT}H,I). The marked growth of $2$-dimensional cavities could reflect an alternative drive to expand and complexify knowledge networks. All empirical Betti curves are significantly different from the Betti curves for the null model data ($p_{perm} < 0.001$) as determined via permutation testing. In summary, across both individual and collective knowledge networks, our findings suggest that information gap theory explains how separate areas of interest ($0$-cycles) grow and then are subsequently linked together, and how loop-like holes ($1$-cycles) within specific areas of interest grow and are subsequently filled. However, the extent and longevity of larger pocket-like voids ($2$-cycles) remains unexplained by the information gap theory, motivating an assessment of alternative psychological drives.

\subsection{Compression progress theory}

\begin{figure}[htbp] 
    \begin{center}
    \includegraphics[width=\textwidth]{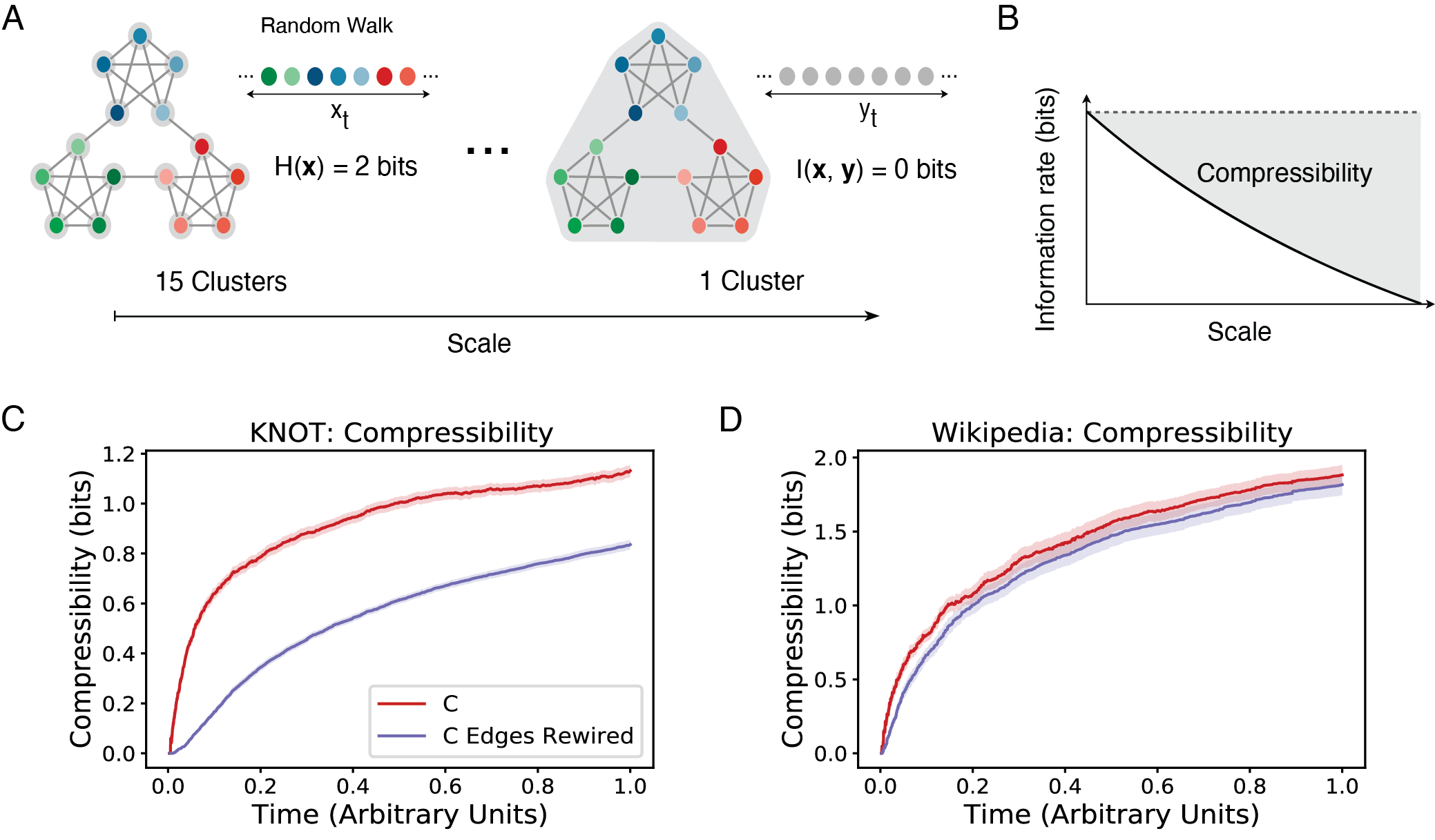}
    \end{center}
    \caption{\label{fig:CPT} \textbf{Quantifying compression progress using network compressibility.} \emph{(A)} A random walk $\boldsymbol{x}$ on a network is a sequence of nodes constructed by transitioning from a node $x_t$ to one of its neighbors uniformly at random. Such a sequence generates information at a rate given by its \textit{entropy} $H(\boldsymbol{x})$. Now suppose that we group the nodes into different clusters; the number of clusters defines the scale at which the network is described. The random walk $\bm{x}$ is compressed into a new sequence $\boldsymbol{y}$, where $y_t$ is the cluster that contains node $x_t$. The clustered sequence $\boldsymbol{y}$ generates information about the original sequence $\bm{x}$ at a rate given by the \textit{mutual information} $I(\boldsymbol{x}, \boldsymbol{y}) = H(\boldsymbol{y}) - H(\boldsymbol{y} | \boldsymbol{x})$. Mutual information $I(\boldsymbol{x}, \boldsymbol{y})$ is greatest---and equal to the entropy $H(\boldsymbol{x})$---when each node is assigned independently to its own cluster. By contrast, in the limit where the entire network is viewed as one large cluster, the mutual information vanishes. \emph{(B)} At each intermediate scale between these two extremes, we can find an optimal clustering that maximally lowers the information rate. Network compressibility is then defined as the maximal reduction in the information rate, averaged across all scales. \emph{(C)} Growing individual knowledge networks are markedly more compressible than expected considering related edge-rewired null model networks. \emph{(D)} Growing collective knowledge networks show only a slight tendency for greater-than-expected compressibility. Shaded regions in panels \emph{C} and \emph{D} represent standard error. Purple curves denote average compressibility values for edge-rewired null model networks.}
\end{figure}

Originally proposed as a general algorithmic framework for reinforcement learning, compression progress theory posits that curiosity is the drive to continually improve the compression of a learner's mental model of the world \citep{Schmidhuber_2008}. By conceptualizing mental models as knowledge networks, we can measure compressibility using recent advances at the intersection of information theory and network science \citep{Lynn_2021}. To compute the compressibility of a network, we begin by considering a random walk $\boldsymbol{x} = (x_1, x_2, \cdots)$, where $x_t$ is the node that appears at step $t$ (Fig. \ref{fig:CPT}A). The rate at which the sequence $\bm{x}$ generates information is given by its entropy $H(\boldsymbol{x})$. If we group the network's nodes into clusters, we can re-write $\boldsymbol{x} = (x_1, x_2, \cdots)$ as $\boldsymbol{y} = (y_1, y_2, \cdots)$ by replacing each node $x_t$ with its cluster identity $y_t$. The rate at which the clustered sequence $\bm{y}$ generates information about the original sequence $\bm{x}$ is given by the mutual information $I(\boldsymbol{x}, \boldsymbol{y}) = H(\boldsymbol{y}) - H(\boldsymbol{y} | \boldsymbol{x})$. The number of clusters that we use to compress the network defines a scale of its description. As we decrease the number of clusters---that is, as we increase the scale of description---the information rate $I(\bm{x}, \bm{y})$ decreases. When each node belongs to its own cluster, the information rate $I(\bm{x},\bm{y})$ equals the original rate $H(\bm{x})$ (Fig. \ref{fig:CPT}A). By contrast, when all nodes are grouped together into one cluster, the information rate is zero (Fig. \ref{fig:CPT}A). At all scales in between, we can find the optimal clustering of nodes that minimizes the information rate (Fig. \ref{fig:CPT}B). We then define the compressibility of the network as the maximal reduction in the information rate, averaged across all scales of its description (Fig. \ref{fig:CPT}B) \citep{Lynn_2021}. 

Considering compression progress theory, we hypothesized that growing knowledge networks would be more compressible than topologically similar edge-rewired null model networks. We test our hypothesis by computing network compressibility for each subgraph in filtrations of individual and collective knowledge networks. We find that compressibility increases monotonically as knowledge networks grow. At all stages of growth, and in support of our hypothesis, networks for individuals exhibit greater-than-expected compressibility (Fig. \ref{fig:CPT}C). This same trend holds, but to a much weaker extent in the collective knowledge networks. While the early stages of growth evince greater separation between empirical and null compressibility values, the two curves overlap in later stages of growth (Fig. \ref{fig:CPT}D). Based on non-parametric permutation testing, compressibility curves for individual and collective knowledge networks are significantly different from their null model counterparts ($p_{perm} < 0.001$). These data provide evidence that is critical for an evaluation of compression progress theory. Consistent with the theory, for individuals, a preference for greater compressibility indicates that curiosity is driven to construct parsimonious mental representations of knowledge. For collectives, a similar-to-expected compressibility could reflect (i) the group's nature as constituted by the diverse voices and expertise that comprise it, which can enhance the relevance of details and preclude their compression, and (ii) the fact that groups may not be constrained by the same cognitive capacity limitations that constrain individuals.

\subsection{Conformational change theory}

A curious learner practising curiosity solely according to information gap theory strives for growth and completeness of knowledge. By contrast, a learner practising curiosity solely according to compression progress theory strives to uncover the latent organization of the world. In the process, neither individual can keep pace with the growing complexity of the environment; with a rapidly expanding frontier of ignorance as new unknowns become accessible. Crucially, both theories suggest how we can usefully add or relinquish information but neither acknowledges the worth of what we already possess. Prior work has shown that curiosity-driven information acquisition is not only about growing or shedding knowledge, but also about retreading and reconsidering what one presently holds \cite{Zhou_2020, LydonStaley_2021}. Following \citet{Zurn_2021}, we propose that such reflection entails moving concepts flexibly in relation to one other. Specifically, we define curiosity as the process of constructing knowledge networks with a finely arbitrated balance between local internal rigidity and global external flexibility. Rigidity and flexibility are mechanical notions that require an object of interest to be embedded in physical space. Therefore, drawing inspiration from a rich literature on cognitive maps (see Supplement for background), we assume that knowledge networks are embedded in Euclidean space where they possess several degrees of freedom. We then measure flexibility as a network's ability to undergo conformational changes \citep{Kim_2019} and formalize our account as the \textit{conformational change theory} of curiosity.

\begin{figure}[htbp] 
    \begin{center}
    \includegraphics[width=\textwidth]{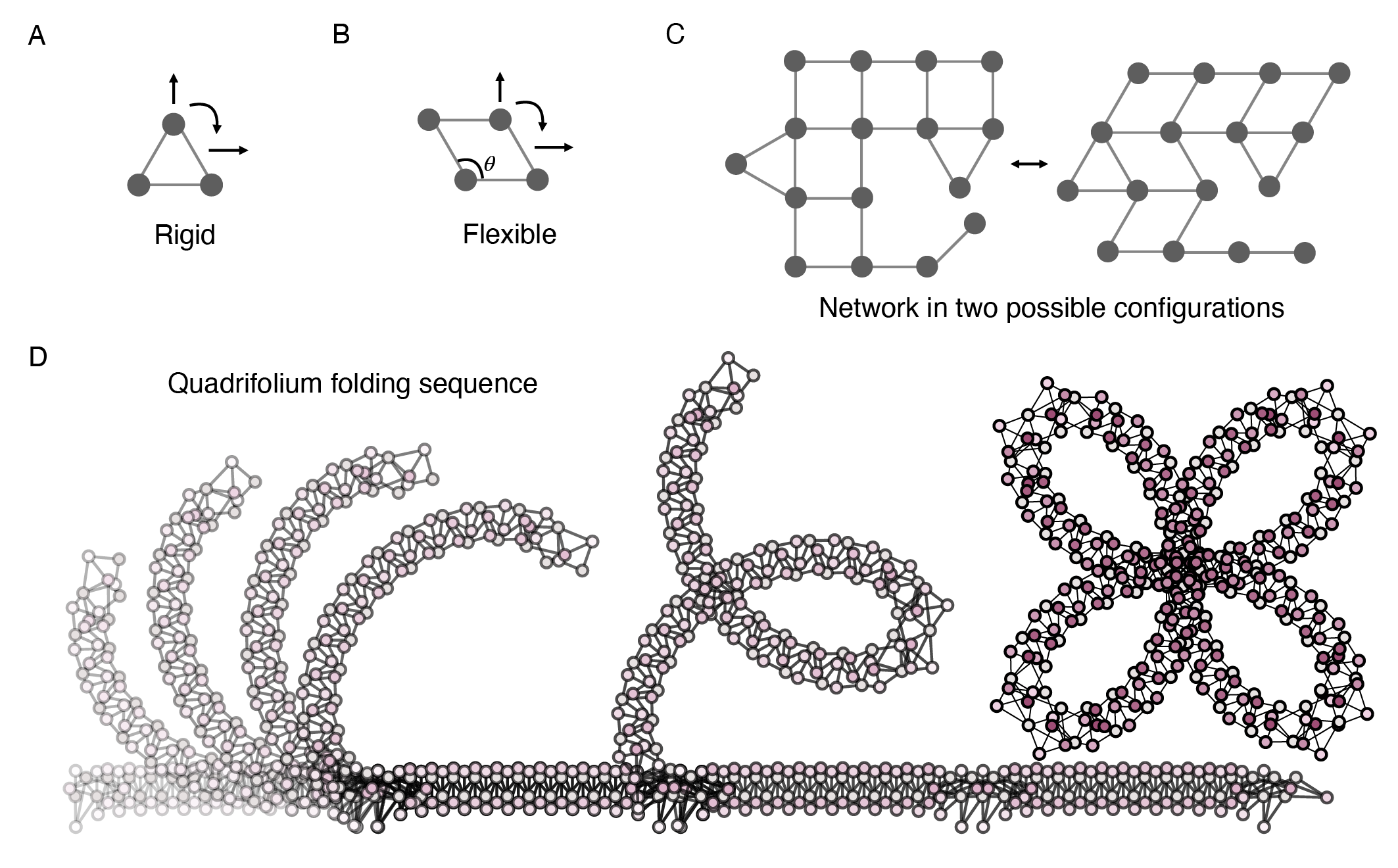}
    \end{center}
    \caption{\label{fig:cct_1} \textbf{Conformational change in mechanical networks.} \emph{(A)} In two-dimensional space, a network with three nodes and three edges has three rigid-body degrees of freedom: horizontal translation, vertical translation, and rotation. \emph{(B)} In addition to the three rigid-body motions, a quadrilateral frame also possesses a \textit{conformational degree of freedom}, depicted here via the angle parameter $\theta$, which allows it to change shape from a square to a diamond. \emph{(C)} Rigid and flexible sub-units can be combined to construct networks capable of undergoing large-scale conformational changes. Different configurations of the same network can be reached by propagating conformational changes through its structure. \emph{(D)} A network chain with $338$ nodes and $672$ edges folds to form a \textit{quadrifolium} (panel \emph{D} reproduced with permission from \citet{Kim_2022}).}
\end{figure}

Before measuring the conformational flexibility of growing knowledge networks, we offer a brief introduction to mechanical networks. Consider a triangular network in two dimensions. Each of its nodes can be located with two coordinates (Fig. \ref{fig:cct_1}A). This network has three available rigid body motions: horizontal translation, vertical translation, and rotation. Next, consider a network comprised of $4$ nodes and $4$ edges (Fig. \ref{fig:cct_1}B). This network possesses the same rigid-body motions as are available to the triangle. Additionally, the quadrilateral possesses a conformational degree of freedom. A conformational change in a network alters the Euclidean distance between unconnected pairs of nodes. For instance, if a pair of adjacent nodes in the quadrilateral is held fixed in space, the remaining nodes can be moved freely while sweeping across an angle $\theta$ with respect to the fixed pair (Fig. \ref{fig:cct_1}B). Through this process, this simple network exhibits a \textit{conformational change} from a square to a diamond. Mechanical networks can exist in several configurations, each of which can be reached through a series of conformational changes from any of the others (Fig. \ref{fig:cct_1}C-D). The number of independent conformational motions available to a network with $p$ nodes and $q$ edges embedded in a $d$-dimensional space is $dp - q$. Among these $dp - q$ degrees of freedom are ${d(d+1)}/{2}$ rigid body motions, which include translations and rotations. The rest, given by
\begin{equation} \label{eq:DoF_C}
    DoF_C = dp-q - \frac{d(d+1)}{2},
\end{equation}
are the available conformational degrees of freedom, or \emph{conformational motions}. 

\begin{figure}[htbp] 
    \begin{center}
    \includegraphics[width=\textwidth]{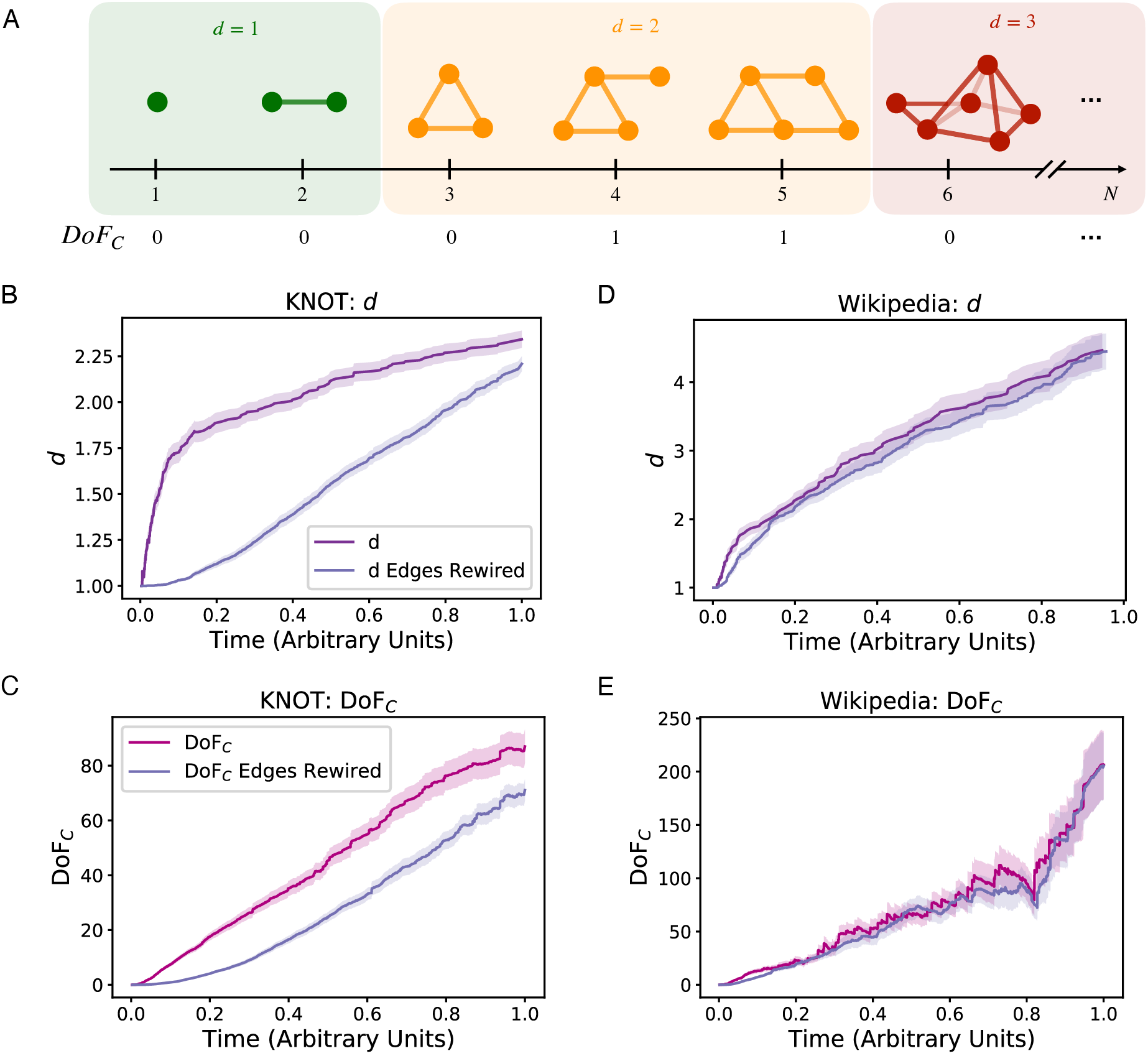}
    \end{center}
    \caption{\label{fig:cct_2} \textbf{Conformational change theory of curiosity.} We propose that in the networked space of the mind, while some concepts and their relationships have fixed locations, others can move flexibly in a context-dependent manner. Such flexibility affords curious humans the ability to rethink and reconfigure what they already know in light of new information. We formalize flexibility as the number of conformational degrees of freedom ($DoF_C$). In a network in $d$-dimensional space with $p$ nodes and $q$ edges, $DoF_C = dp-q - d(d+1)/2$. Assuming $d = 1$ initially, we compute $DoF_C$ for filtrations of growing knowledge networks. A negative value for $DoF_C$ indicates the presence of self-stress, which we resolve by incrementing the $d$ by $1$. \emph{(A)} In the example filtration, when nodes $3$ and $6$ are added, the network becomes over-constrained and develops self-stress. Consequently, dimensionality first increases from $1$ to $2$ and then from $2$ to $3$. \emph{(B, C)} Individual knowledge networks require greater dimensionality and possess greater flexibility than null model networks. \emph{(D, E)} Collective knowledge networks do not exhibit greater dimensionality or conformational flexibility than null model networks. Shaded regions in \emph{B}-\emph{E} represent the standard error.}
\end{figure}

Crucially, Eq. \ref{eq:DoF_C} relies on the linear independence of edges. Linear independence entails that there are no redundant edges that over-constrain a set of nodes beyond the formation of a rigid cluster yielding a state of self-stress, and that the network does not exist in a rare and pathological geometry known as a kinematic bifurcation \cite{Mao2018Topological, Kim_2019}. States of self-stress imply that edges within a network bear internally balanced forces. A negative value for the number of conformational degrees of freedom would indicate that the network---when considered in its entirety---is over-constrained. In our framework, we assume that such states result in a form of \textit{cognitive dissonance} whereby competing constraints between concepts cannot be resolved. We alleviate this tension by incrementing the dimensionality by $1$ when needed. Specifically, we increment $d$ by $1$ until $DoF_C$ is no longer negative. This approach yields the minimum dimensionality required at each stage of growth to avoid over-constraining the network. Fig. \ref{fig:cct_2}A depicts this process for a representative filtration of a growing network. When node $3$ is added to the network, which is initially embedded in a $1$-dimensional space (green), the number of conformational degrees of freedom evaluates to $(1 \times 3)-(3) - 1 = -1$, indicating the presence of self-stress and requiring that the embedding dimensionality be incremented to $2$ (orange). This process repeats when node $6$ is added, resulting in an increase in dimensionality to $3$ (red). To compute the number of conformational degrees of freedom for growing knowledge networks, we assume that they are initially embedded in a $1$-dimensional space; whenever the quantity in Eq. \ref{eq:DoF_C} becomes negative, we increment dimensionality by $1$.

We hypothesized that knowledge networks would possess greater conformational flexibility than corresponding null model networks. We test this hypothesis by computing the number of conformational degrees of freedom in filtrations of individual and collective knowledge networks (Fig. \ref{fig:cct_2}C, E). In parallel, we track the minimum embedding dimensionality required to prevent self-stress from developing in the growing networks (Fig. \ref{fig:cct_2}B, D). We find that individual knowledge networks need greater dimensionality and possess greater conformational flexibility than null model networks (Fig. \ref{fig:cct_2}B, C). By contrast, measurements of dimensionality and flexibility for collective networks cannot be as easily distinguished from their corresponding null model data (Fig. \ref{fig:cct_2}D, E). However, for both data sets, the empirical curves for dimensionality and conformational flexibility are significantly different from the corresponding curves from the null model data ($p_{perm} < 0.001$). Our findings suggest that individuals value the ability to reconsider what they already know in light of newly acquired information. On the other hand, collective knowledge displays less capacity for global reconfiguration over the long time scales evaluated in this study; future work could investigate the existence and dynamics of internal sectors that change shape over different time scales or during paradigm shifts.

\section{Discussion}

In this work, we formalize curiosity as the process of constructing a growing knowledge network. We leverage tools from network science to quantitatively examine several theoretical constructs for curiosity such as the information gap theory and compression progress theory. Information gap theory suggests that curiosity is the drive to obtain units of knowledge that fill gaps in understanding \citep{Loewenstein_1994}. Compression progress theory posits that curiosity is the drive to uncover the latent organization of the world \citep{Schmidhuber_2008}. We probe information gaps as topological cavities in growing knowledge networks and quantify compression progress using network compressibility. The two theories offer complementary perspectives on curiosity; the information gap theory suggests that new information is acquired to fill knowledge gaps, whereas the compression progress theory suggests that new information is used to distill the essential epistemic elements of knowledge. While these perspectives describe how knowledge networks become denser and simpler through information acquisition, an alternative formulation is needed to explain how they become expansive and more complex. Therefore, we build upon a recently proposed conceptual framework \citep{Zurn_2021} to develop the conformational change theory of curiosity. We posit that knowledge networks are embedded in a Euclidean geometry, which allows concepts to move flexibly in relation to one another. We then view curiosity as the practice of constructing mechanically flexible knowledge networks. Formally, we measure conceptual flexibility as the number of conformational degrees of freedom available to a growing knowledge network. Throughout our investigations, we take a multi-scale view and probe evidence for each theory in individuals and in collectives. Across the two scales of granularity, we determine the precise contexts in which each theoretical account is explanatory, thereby clarifying their complementary and specific affordances.

\textbf{Information gap theory and topological cavities in knowledge networks.} Information gap theory suggests that humans tolerate a finite amount of uncertainty in their knowledge of the world \citep{Loewenstein_1994}. Exposure to a small amount of previously unknown information brings into focus the presence of a knowledge gap, pushing the level of uncertainty past an acceptable threshold. This increased uncertainty then prompts a search for information to fill the knowledge gap and resolve the unknown. In this work, we formalize gaps as topological cavities in growing knowledge networks and track their evolution in dimensions $0$, $1$, and $2$ \citep{Sizemore_2018_1, Ju_2020, Bianconi_2021}. Each dimension is characterized by a different kind of topological gap: $0$-dimensional gaps correspond to disconnected network components, $1$-dimensional gaps correspond to loop-like holes, and $2$-dimensional gaps correspond to pocket-like voids. Across all dimensions, we find that the number of cavities increases as individual knowledge networks grow. Stated differently, associations between familiar concepts remain undiscovered even as we acquire more information. Hence, in addition to the common view of an expanding frontier of ignorance, knowledge growth is accompanied by an ever-expanding interior of ignorance \cite{Ju_2020}. Except for the $0$-th dimension, we report similar results for knowledge networks built collectively. Filling a $0$-dimensional cavity entails adding an edge between two disconnected network components. Such edges may be easier for collectives to add than for individuals since interdisciplinary sub-fields within scientific domains are motivated to link disparate sub-areas of knowledge \citep{Keisuke_2019}. Importantly, and in support of the information gap theory, the number of $0$- and $1$-dimensional cavities is lower in observed individual and collective knowledge networks than in the corresponding null model data, reflecting a downward pressure on the number of gaps created, consistent with a gap-filling drive. Therefore, from a networks perspective, gaps---as envisioned by information gap theory, those that are prioritized for filling---may best correspond to topological cavities of dimensions $0$ and $1$. Stated differently, information gap theory provides an explanation for the markedly damped growth of lower dimensional cavities; however, a different account is needed to explain the contrasting proliferation of higher dimensional cavities, both in individuals and in collectives.

\textbf{Compression progress theory and efficient network representations of knowledge.} To gain a deeper intuition, we turn to compression progress theory, which derives inspiration from resource limitations that underpin brain function \citep{Schmidhuber_2008}. We represent knowledge as a network of concepts and their inter-relationships, and we compute network compressibility \citep{Lynn_2021} to determine whether curiosity drives compression. We find that growing individual knowledge networks consistently exhibit greater-than-expected compressibility, consistent with the theory. This finding can be contextualized by considering the fact that as we interact with the world, we encounter and consume large quantities of information. Constructing perfectly accurate mental models would entail storing each unit of acquired knowledge separately. However, finite resources constrain us to build compressed or efficient abstractions of observed data that can generalize across contexts \citep{Tenenbaum_2011}. According to compression progress theory, information that---when acquired---facilitates such abstraction is more valuable \citep{Schmidhuber_2008}. Our results support this proposition and suggest that individuals preferentially seek such information. By contrast, the compressibility curve for collective knowledge networks tends to align with the curve for the corresponding null model data in later stages of growth. This finding can be contextualized by considering the fact that collectives can store vast quantities of detailed information in a distributed manner and, hence, do not face the same resource limitations that individuals do. In summary, while compression progress theory is supported by our data from individual knowledge networks, the building of collective knowledge networks appears to require a different account.


\textbf{Conformational change theory and the mechanical flexibility of knowledge networks.} The conformational change theory of curiosity is an alternative account that is built on two assumptions. First, we assume that humans encode conceptual knowledge in cognitive networks. Second, we assume that knowledge networks are embedded in Euclidean space, where they possess several degrees of freedom. Both assumptions are predicated on how humans encode spatial and abstract knowledge \citep{Garvert_2017, Warren_2019, Peer_2021, Stiso_2022}. Evidence from spatial navigation studies demonstrates that mental representations of space take the form of labeled cognitive graphs. Each node represents a physical location and is accompanied by local metric information such as angles and Euclidean distances to its immediate neighbors \citep{Chrastil_2014, Warren_2019, Peer_2021}. Furthermore, hexadirectional modulation, which is the telltale signature associated with an underlying map-like neural code, is observed in the neural signals when individuals navigate discrete and continuous abstract concept spaces \citep{Constantinescu_2016, Park_2021} (see Supplement for details on mental representations of spatial and non-spatial knowledge). Building on Euclidean cognitive graphs, we operationalize conceptual flexibility in knowledge networks as the number of conformational degrees of freedom. We find that growing individual knowledge networks have greater-than-expected embedding dimensionality and conformational flexibility. According to conformational change theory, embedding dimensionality increments when growing knowledge networks become over-constrained and develop self-stress. We find that such stress arises more frequently in individual knowledge networks than in null model data. This observation is consistent with the conformational change theory of curiosity, and suggests that individuals' idiosyncratic acquisition of information leads to a frequent reshaping of concept relations based on context. By contrast, in knowledge networks built collectively we find that the evolution of mechanical features-of-interest cannot be distinguished from their evolution in null model data. Collective networks grow through a dynamic interplay of consensus and dissensus between large groups of individuals. Therefore, it is possible that due to the long time scales that we focus on in this study, dynamic events associated with collective knowledge growth, such as paradigm shifts, are simply concealed from view in local sectors of each field.



\textbf{Using computational measures to operationalize conceptual theories.} Our study of curiosity theories stands on the backdrop of significant work in psychology that delineates different types of information acquisition. Classical perspectives from psychology separate notions of curiosity into two broad categories \citep{Berlyne_1954}. \textit{Perceptual} curiosity is the desire for increased perception of specific sensory stimuli and is reduced via continued exposure to such stimuli. By contrast, \textit{epistemic} curiosity is a more general desire for non-perceptual knowledge. Based on the breadth of knowledge sought, epistemic curiosity is further classified as \textit{specific} or \textit{diversive} \citep{Berlyne_1960}. Specific curiosity is associated with the desire to reduce uncertainty about an ambiguous stimulus and leads to exploration in search of a particular piece of information. Diversive curiosity is less restrictive and refers to the desire to obtain wide-ranging knowledge. Each category represents a different manifestation of self-driven information-seeking behavior. Across these sub-types, many scales have been developed to quantify curiosity in individuals \citep{Collins_2004, Reio_2006, Litman_2008, Kashdan_2018, Wagstaff_2021}. However, such efforts rely mainly on questionnaires to measure curiosity at discrete points in time and are unable to record its dynamic nature. By formulating curiosity as a process of knowledge network building, we offer calculable, theory-based measures such as topological gaps, compressibility, and conformational degrees of freedom that can be used to characterize the longitudinal process of information-seeking, whether in laboratory experiments or in the wild. Important future directions include mapping each network measure to the most suitable sub-type of curiosity and expanding the current taxonomy of sub-types to accommodate novel network theoretical perspectives.

\textbf{Implications for the study of reinforcement learning.} The computational metrics we examine are relevant not only for the study of human curiosity, but also potentially for that of artificial intelligence. Compressibility, for instance, was originally proposed as an intrinsic learning signal to guide reinforcement learning \citep{Schmidhuber_2008}. Our work provides several candidate metrics---such as the number of topological cavities, network compressibility, and conformational flexibility---that can act as suitable curiosity-based signals for task settings where the environment can be modeled as a network. Out of the three measures, while it is currently computationally intensive to compute persistent homology and network compressibility, the number of conformational degrees of freedom can be determined inexpensively. Information acquisition in reinforcement learning is a means to an end, where the end is a reward associated with the successful completion of a specific task \citep{Gottlieb_2013, Sutton_2018}. An agent seeking to collect high total reward during interactions with its environment must strike a balance between \textit{exploitation} and \textit{exploration}. The agent must exploit, or productively use, those actions that are currently known to yield high reward but must also occasionally explore untested actions that may eventually turn out to be better. In many real-world settings, external rewards are highly infrequent or even completely absent and, thus, cannot reliably guide behavior. In such sparse reward environments, curiosity-like intrinsic motivations can lead to improved exploration and, by extension, improved task performance \citep{Pathak_2017, Savinov_2018}. The design of intrinsic (or curiosity-based) reward signals for reinforcement learning is an increasingly important area for further research \citep{Aubret_2019}, and may benefit from computational insights into human behavior, such as those derived from our analyses here.

\noindent \textbf{Methodological considerations.} Several methodological considerations are pertinent to our work. First, in the applied mathematics literature, a network filtration is typically specified based on rank-ordered edges as opposed to rank-ordered nodes \citep{Giusti_2015, Giusti_2016, Sizemore_2018_2}. Edges are first ordered by decreasing weights. The filtration then proceeds by sequentially incorporating higher-ranked edges into the network. Here instead, we assign ranks to nodes and construct node-ordered filtrations of knowledge networks. At each growth stage, we add a node and all edges that exist between the node and other nodes already present in the network. It is possible that future efforts could gain greater granularity in the constructed filtrations by assigning ranks to both nodes and edges. For instance, edges (hyperlinks) that individuals click may be ranked higher than edges that individuals do not click. Second, we construct knowledge networks for individuals based on Wikipedia browsing data acquired for fifteen minutes each day for three weeks. However, in reality daily knowledge acquisition is continuous and not limited to Wikipedia as a source. Furthermore, participants may act on their curiosity about a specific concept during un-tracked sessions and continue to explore related ideas during tracked sessions. The KNOT study, where the data for individuals are sourced from, implicitly mitigates such discontinuities by linking the last page of each browsing session to the first page of the following session. Third, in the conformational change theory of curiosity, we resolve states of self-stress by incrementing $d$ by one until $DoF_C$ is no longer negative. Importantly, we only consider self-stress in the entire ensemble of nodes and edges. It is possible, however, that certain subsets of the network may be over-constrained, even if the entire network, as a whole, may not appear to be. Such over-constrained subgraphs can be identified using algorithms such as the pebble game \citep{jacobs1995generic}. Self-stress may also occur due to pathological geometries in networks referred to as kinematic bifurcations \citep{Kim_2019}. To address such geometries, we require a precise embedding of concepts in Euclidean space, which remains an important area for future work.

\section{Conclusion}
We conceptualize curiosity as a process of knowledge network building in order to examine three theoretical accounts: information gap theory, compression progress theory, and conformational change theory. Formalizing curiosity in terms of networks helps us to \emph{quantitatively} operationalize predominantly \emph{qualitative} theoretical constructs. Information gaps can be identified as topological cavities, compression progress can be quantified using network compressibility, and flexibility---as premised on the conformational change theory---can be quantified as the number of conformational degrees of freedom. We use data acquired from Wikipedia to construct growing knowledge networks for individuals and for collectives. We find that as networks grow, knowledge gaps increase in number, suggesting an expanding interior of ignorance. Yet, in support of an aversion to gaps predicted by information gap theory, we also find fewer-than-expected disconnected network components (or $0$-dimensional topological cavities) and fewer-than-expected loops of edges (or $1$-dimensional topological cavities) in growing knowledge networks. This set of findings suggests that knowledge ``gaps'' as conceptualized by information gap theory may best translate, in a network theoretical sense, to $0$ and $1$-dimensional cavities. We also find that growing individual knowledge networks possess greater-than-expected compressibility, indicating that information acquisition is driven to construct parsimonious mental world models. In addition, we find that knowledge networks built by individuals become increasingly flexible with growth, foregrounding the longstanding relevance of conformational change in the mind. Our results lend support to a pluralistic view of curiosity, wherein intrinsically motivated information acquisition fills knowledge gaps and builds increasingly compressible and flexible mental representations of the world. Our findings offer a novel network theoretical perspective on intrinsically motivated information acquisition that may harmonize with or compel an expansion of the classical taxonomy of curiosity. 
 
\section{Methods}

\subsection{Data}

\subsubsection{Knowledge networks built by individuals}

Knowledge networks for individuals are constructed with data obtained from the ``Knowledge Networks Over Time'' (KNOT) study \citep{LydonStaley_2020_1, LydonStaley_2020_2, LydonStaley_2021}. These data are comprised of Wikipedia browsing histories of 149 individuals (121 women, 26 men, 2 other) collected between October 2017 and July 2018. At the time of data acquisition, participants were aged between $18.21$ and $65.24$ years ($\mu = 25.05$, $\sigma = 6.99)$; $6.71$\% identified as African American/Black, $25.50$\% identified as Asian, $5.37$\% identified as Hispanic/Latino, $49.66$\% identified as White, $5.37$\% identified as Multiracial, $5.37$\% identified as Other, and $2.01$\% provided no racial or ethnic information. Every evening for $21$ days, participants were prompted to open a browser and navigate to \texttt{wikipedia.org}. They were then instructed to engage in $15$ minutes of self-directed information search with no restrictions placed on how they could traverse from one page to another. At the end of each session, participants used tracking software (\texttt{Timing}), pre-installed on their personal computers or laptops, to export and upload their browsing histories. 

We treat all pages visited by an individual as nodes in a knowledge network. Edges between nodes are specified based on the presence of hyperlinks. Prior work has found that pairs of pages connected by hyperlinks are significantly more similar to each other compared to pairs that are not connected by hyperlinks \citep{LydonStaley_2021}. Thus, we add an undirected and unweighted edge between \texttt{Page 1} and \texttt{Page 2} if either \texttt{Page 1} links to \texttt{Page 2} or \texttt{Page 2} links to \texttt{Page 1}. Hyperlinks are not required to exist bidirectionally for an edge to exist between two nodes. We determine the presence of hyperlinks based on how Wikipedia appeared on August 1, 2019 and adapt code from the \texttt{wikinet} package for network construction \citep{Ju_2020}. Each node (or page) in the browsing data is accompanied by an index that denotes the temporal order in which it was visited. For every individual, the nodes and edges as well as the order of node visitation is used to specify a growing knowledge network.

\subsubsection{Knowledge networks built collectively}

In its role as an encyclopedia, Wikipedia represents a large repository of knowledge acquired over thousands of years through collective human effort. Building on prior work, we construct domain-specific collective knowledge networks by taking subgraphs of the larger Wikipedia network \citep{Ju_2020}. Information in Wikipedia is organized in a hierarchical manner, which makes it possible to identify articles that pertain to a particular domain of interest. We capitalize on this structure to construct knowledge networks for thirty topics: abstract algebra, accounting, biophysics, Boolean algebra, calculus, cognitive science, commutative algebra, dynamical systems and differential equations, dynamical systems, earth science, economics, education, energy, evolutionary biology, geology, geometry, group theory, immunology, linear algebra, linguistics, meteorology, molecular biology, number theory, optics, philosophy of language, philosophy of law, philosophy of mind, philosophy of science, sociology, and software engineering. All pages listed under a topic are treated as nodes in the topic's network. For instance, the network for molecular biology contains pages for `allele', `lymphocyte', and `antibody' as nodes. Similar to knowledge networks for individuals, edges between nodes are added on the basis of hyperlinks. Typically, articles also contain information about the year in which the concept they describe first became known; the year attribute is used as an index to specify node order in a growing graph. More details on the network construction process (such as the procedure followed when a page has no year attribute) are available from Ref. \citet{Ju_2020}.

\subsection{Detecting topological cavities} \label{sec:homology}

In order to identify cavities of various dimensions in a network, we construct a higher-order relational object known as a \textit{simplicial complex}. While a graph is comprised of a set of nodes and a set of edges, a simplicial complex consists of simplices. A \textit{simplex} represents a polyadic relationship among a finite set of $k$ nodes. Geometrically, a $k$-simplex is realized as the convex hull (enclosure) of $k+1$ generally placed vertices. For $0 \leq k \leq 2$, a node is a $0$-simplex, an edge is a $1$-simplex, and a filled triangle is a $2$-simplex. Simplices follow the \textit{downward closure} principle, which requires that any subset of vertices, known as a \textit{face}, within a simplex also form a simplex. For instance, a $2$-simplex (filled triangle) has three $1$-simplices (edges) as its faces, each of which in turn is comprised of two $0$-simplices (nodes). In graph theoretical terms, a $k$-simplex corresponds to a $(k+1)$-clique, which is an all-to-all connected subgraph of $k+1$ nodes. We can construct a simplicial complex by assigning a $k$-simplex to each $(k+1)$-clique in a binary graph. Thus, the resulting combinatorial object is sometimes also referred to as the \textit{clique complex} of the graph. We denote the clique complex of the graph $\mathcal{G}_p$ as $\mathcal{X}(\mathcal{G}_p)$.

In a clique complex, a $k$-dimensional topological cavity is identified as an empty enclosure formed by $k$-simplices. Whether a collection of simplices encloses a cavity is determined in part by its \textit{boundary}. The boundary of a $k$-simplex $\sigma$ is defined as the set $\partial \sigma$ of its $(k-1)$-faces. The boundary of a set of simplices $K = \{\sigma_1,\;\sigma_2,\;\cdots,\;\sigma_m\}$ is obtained by taking the symmetric difference $\Delta$ of the boundaries of its constituents,
\begin{equation*}
    \partial K = \partial \{\sigma_1,\;\sigma_2,\;\cdots,\;\sigma_m\} = \partial \sigma_1 \; \Delta \; \partial \sigma_2 \; \Delta \; \cdots \; \Delta \; \partial \sigma_m.
\end{equation*}
The symmetric difference is an associative operation that returns the union of two sets without their intersection. A set of $k$-simplices with an empty boundary is called a \textit{k-cycle}. At first glance, it may seem adequate to identify cycles of various dimensions in a simplicial complex and treat them as topological cavities. However, note that any collection of $(k+1)$-simplices has a $k$-cycle as its boundary. For example, the boundary of a $2$-simplex is a $1$-cycle that is ``filled in'' by the $2$-simplex. Thus, it is necessary to distinguish non-trivial cycles that constitute true cavities from those that trivially belong to the boundaries of higher-dimensional simplices. Finally, we introduce the notion of \textit{equivalence}. Two $k$-cycles $K_1$ and $K_2$ are equivalent if $K_1\;\Delta\;K_2$ is the boundary of a collection of $(k+1)$-simplices. \textit{Homology} refers to the counting of non-equivalent cycles of various dimensions in a clique complex. It is customary to refer to non-equivalent cycles simply as \emph{cycles} for brevity.

The graph filtration from Eq. \ref{eq:graph_filtration} induces a related filtration of clique complexes
\begin{equation} \label{eq:clique_complex_filtration}
    \mathcal{X}(\mathcal{G}_0) \subset \mathcal{X}(\mathcal{G}_1) \subset \cdots \subset \mathcal{X}(\mathcal{G}_N) = \mathcal{X}(\mathcal{G}).
\end{equation}
At each stage in the filtration, we add a node and replace all cliques that may result from its addition with relevant simplices. While some newly added simplices create cavities, others close older ones. Equivalence allows us to compute \textit{persistent homology} wherein we track the evolution of each cavity from the moment it is first born to the moment it is completely filled in by higher simplices. At any index $p$ of the filtration, the $k$-th \textit{Betti number} $\beta_k(p)$ records the number of active cavities of dimension $k$. We then define the $k$-th Betti curve as the sequence of numbers $\{\beta_k(p)\}_{p = 0}^{N}$. We compute persistent homology for all knowledge networks using the \texttt{Ripser.py} package in \texttt{Python} \citep{Tralie_2018}. Visualizations of persistent homology results are generated with code adapted from \citep{Christianson_2020}. 

For a more comprehensive treatment of topological data analysis, we direct the interested reader to Refs. \citep{Carlsson_2009, Ghrist_2007, Hatcher_2002, Zomorodian_2005, Sizemore_2019, Bianconi_2021}.

\subsection{Computing network compressibility}

In order to estimate the compressibility of a network, we consider a binary graph $\mathcal{G}_p$ with $p$ nodes and $q$ edges, which can be represented by a symmetric adjacency matrix $M \in \R^{p \times p}$. A message containing information about the network's structure can be conveyed to an arbitrary receiver by encoding it in the form of a random walk $\boldsymbol{x} = \left(x_{1},\; x_{2},\; \ldots\;\right)$. The walk sequence is generated by transitioning from a node to one of its neighbors uniformly at random. Thus, for a random walk on $\mathcal{G}_p$, the probability of transitioning from node $i$ to node $j$ is $P_{ij} = M_{ij}/\sum_jM_{ij}$. Since the random walk is Markovian, the rate at which such a message transmits information (or its \textit{entropy}) is given by
\begin{equation} \label{eq:random_walk_entropy}
    H(\boldsymbol{x})=-\sum_{i} \pi_{i} \sum_{j} P_{i j} \log P_{i j}.
\end{equation}
Here, $\pi_i$ is the stationary distribution representing the long-time probability that the walk arrives at node $i$, which is given by $\pi_i = \sum_jM_{ij}/2q$.

Assigning clusters to nodes leads to a coarse-grained sequence $\boldsymbol{y} = \left(y_{1},\; y_{2},\; \ldots\;\right)$, where $y_t$ is the cluster containing node $x_t$. The number of clusters $n$ can be used to define a scale of the network's description $S = 1 - \frac{n-1}{p}$. For example, when $n = p$, the network is described at a fine-grained scale $S = 1/p$; by contrast, when $n = 1$, the network is described at the largest possible scale $S = 1$. In general, the distorted sequence $\boldsymbol{y}$ is non-Markovian. However, we can still use Eq. \ref{eq:random_walk_entropy} to find an upper-bound on its information rate. At every scale of description, it is possible to identify a clustering of nodes that minimizes this upper bound. After computing these optimal clusterings across all scales, we arrive at a rate-distortion curve $R(S)$, which represents the minimal upper bound on the information rate as a function of the scale $S$. The compressibility $C$ of the network is then given as the average reduction in $R(S)$ across all scales \citep{Lynn_2021},
\begin{equation} \label{eq:compressibility}
    C = H(\boldsymbol{x}) - \frac{1}{p}\sum_{S} R(S).
\end{equation}
Visually, this quantity represents the total area above the rate-distortion curve and below the entropy of the original random walk $H(\boldsymbol{x})$ (Fig. \ref{eq:compressibility}B). For a graph filtration such as in Eq. \ref{eq:graph_filtration}, we abuse indexing notation and define the compressibility curve as the sequence $\{C(p)\}_{p = 0}^{N}$, where $p$ denotes the number of nodes in subgraph $\mathcal{G}_p$. 

\subsection{Computing mechanical network features}

Consider a set of nodes $\mathcal{V} = \{1,\dotsm,N\}$ embedded in $d$ dimensions. Each node $i\in\mathcal{V}$ is located at a particular coordinate in space $\bm{x}_i \in \mathbb{R}^d$. On its own, this system possesses $dN$ degrees of freedom, as each node is able to move independently in space. If we connect these nodes with edges in the set $\mathcal{E} \subseteq \mathcal{V}\times\mathcal{V}$, then each edge $e_{ij} \in \mathcal{E}$ between node $i$ and node $j$ removes one degree of freedom along the direction of edge extension. Each edge generates a constraint that keeps the distance between the nodes constant, such that
\begin{align*}
(\bm{x}_i - \bm{x}_j)^\top (\bm{x}_i - \bm{x}_j) = \mathrm{constant}.
\end{align*}
To linear order, this constraint can be modified by taking the total derivative of both sides and dividing by 2 to yield
\begin{align}
\label{eq:orth}
(\bm{x}_i-\bm{x}_j)^\top (\mathrm{d}\bm{x}_i-\mathrm{d}\bm{x}_j) = 0
\end{align}
where $\mathrm{d}$ is the differential operator. Intuitively, Eq.~\ref{eq:orth} is simply a dot product between the vector pointing from $\bm{x}_j$ to $\bm{x}_i$, and the node motions. Hence, Eq.~\ref{eq:orth} implies that the nodes must move \textit{perpendicular} to the edge such that the edge does not change length. If we compile all such constraints for every edge in $\mathcal{E}$ then we obtain $E = |\mathcal{E}|$ constraints on the node motions. If these constraints are independent, then the total number of degrees of freedom is reduced to $dN - E$. Among these degrees of freedom, $d(d+1)/2$ are \textit{rigid body motions} that do not change the distance between \textit{any} pair of nodes. Hence, the number of conformational degrees of freedom is given by
\begin{align*}
DoF_{C} = dN - E - \frac{d(d+1)}{2}.
\end{align*}
This line of reasoning was first put forth by \citet{maxwell1864calculation} in 1864. 

Many important extensions of this idea exist. One important extension considers the violation of independent constraints. This violation can occur in several ways. One such way is over-constraining a network. For example, a network of $N = 4$ nodes embedded in $d=2$ dimensions is over-constrained if we place edges between all node pairs, such that $\mathcal{E} = \{(1,2),(1,3),(1,4),(2,3),(2,4),(3,4)\}$. Here, there are $|\mathcal{E}| = 6$ edges, such that $DoF_C = [(2\times4) - 6] - 3 = -1$. For a network to possess negative degrees of freedom, there must exist patterns of edge compressions and tensions that are \textit{load-bearing}, such that there are balanced internal forces held within the edges and experienced by the nodes \citep{Mao2018Topological}. In the conformational change theory of curiosity, we treat such states of self-stress as aversive. Practically, whenever $DoF_C$ becomes negative, we resolve competing constraints by incrementing the embedding dimensionality by $1$.

\subsection{Statistical testing} \label{sec:statistical_testing}

We use non-parametric permutation testing to determine whether feature curves, such as those for compressibility and conformational flexibility, for empirical knowledge networks differ significantly from those for corresponding null model networks \citep{Ramsay_2005}. For a given feature, we first compute the area $A$ between the average curve for the observed data and the average curve for the null model data using numerical integration in \texttt{Python}. We then pool all data together and randomly re-assign each data point to either the empirical data group or the null model data group. Each group results in a pseudo-curve of values for a given feature of interest. We compute the area $A'$ between the pseudo-curves for the two groups and repeat this process for $I = 1000$ iterations. For the group difference between empirical and null model data, we define the p-value $p_{perm}$ as the number of times $A'$ is greater than $A$ divided by the number of iterations $I$.

\subsection{Data and code availability}
All data used in the manuscript are available upon request from the corresponding author. All code used is available at \texttt{https://github.com/spatank/Curiosity}.

\clearpage
\section{Citation diversity statement}

Recent work in a number of scientific fields has identified a bias in citation practices such that papers by women and other minority scholars are under-cited relative to the number of such papers in the field \citep{mitchell2013gendered,dion2018gendered,caplar2017quantitative, maliniak2013gender, Dworkin2020.01.03.894378, bertolero2021racial, wang2021gendered, chatterjee2021gender, fulvio2021imbalance}. Here, we sought to proactively choose references that reflect the diversity of the field in thought, form of contribution, gender, race, ethnicity, and other factors. First, we predicted the gender of the first and last authors of each reference using databases that store the probability of a first name being carried by a woman \citep{Dworkin2020.01.03.894378, zhou_dale_2020_3672110}. By this measure (and excluding self-citations to the first and last authors of our current paper), our references contain 16.27\% woman(first)/woman(last), 11.80\% man/woman, 19.30\% woman/man, 52.63\% man/man citation categorizations. This method is limited in that a) names, pronouns, and social media profiles used to construct the databases may not, in every case, be indicative of gender identity and b) it cannot account for intersex, non-binary, or transgender people. Second, we obtained predicted racial/ethnic category of the first and last author of each reference using databases that store the probability of a first and last name being carried by an author of color \cite{ambekar2009name, sood2018predicting}. By this measure (and excluding self-citations), our references contain 4.67\% author of color/author of color, 9.86\% white author/author of color, 20.34\% author of color/white author, and 65.12\% white author/white author citation categorizations. This method is limited in that a) names, Census entries, and Wikipedia profiles used to make predictions about gender may not be indicative of racial/ethnic identity, and b) it cannot account for Indigenous and mixed-race authors, or those who may face differential biases due to the ambiguous racialization or ethnicization of their names. We look forward to future work that could help us to better understand how to support equitable practices in science.

\section{Acknowledgments}
The authors gratefully acknowledge helpful discussions with Drs. Lorenzo Caciagli, Erin G. Teich, and Kieran Murphy. This work was supported by the Center for Curiosity. The authors would also like to acknowledge additional support from the Army Research Office (Grafton-W911NF-16-1-0474, Falk-W911NF-18-1-0244, DCIST-W911NF-17-2-0181) and the National Institute of Mental Health (1-R21-MH-124121-01). The content is solely the responsibility of the authors and does not necessarily represent the official views of any of the funding agencies.

\clearpage
\bibliographystyle{unsrtnat}
\bibliography{main}
\end{document}